\documentclass[aps,floats]{revtex4}
\usepackage{amsmath,amssymb}
\usepackage{graphicx,epsfig}

\begin{document}
\bibliographystyle {plain}

\def\oppropto{\mathop{\propto}} 
\def\opsimeq{\mathop{\simeq}}
\def\opoverderline{\mathop{\overline}}
\def\operarrow{\mathop{\longrightarrow}}
\def\opsim{\mathop{\sim}} 
\def\opmin{\mathop{\min}} 
\def\opmax{\mathop{\max}} 

\def\fig#1#2{\includegraphics[height=#1]{#2}}
\def\figx#1#2{\includegraphics[width=#1]{#2}}



\title{ Many-Body-Localization : Strong Disorder perturbative approach  \\
 for the Local Integrals of Motion  }


\author{ C\'ecile Monthus }
 \affiliation{Institut de Physique Th\'{e}orique, 
Universit\'e Paris Saclay, CNRS, CEA,
91191 Gif-sur-Yvette, France}
 
\begin{abstract}
For random quantum spin models, the strong disorder perturbative expansion of the Local Integrals of Motion (LIOMs) around the real-spin operators is revisited. The emphasis is on the links with other properties of the Many-Body-Localized phase, in particular the memory in the dynamics of the local magnetizations and the statistics of matrix elements of local operators in the eigenstate basis. 
 Finally, this approach is applied to analyze the Many-Body-Localization transition in a toy model studied previously from the point of view of the entanglement entropy.

\end{abstract}

\maketitle

\section{ Introduction} 

In the field of Many-Body-Localization (see the recent reviews \cite{revue_huse,revue_altman,revue_vasseur,revue_imbrie,revue_rademaker,review_mblergo,review_prelovsek,review_rare} and references therein), the notion of Local Integrals of Motion (LIOMs) has emerged as an essential notion to understand the various properties of the Many-Body-Localized-Phase
\cite{emergent_swingle,emergent_serbyn,emergent_huse,emergent_ent,imbrie,serbyn_quench,emergent_vidal,emergent_ros,
emergent_rademaker,serbyn_powerlawent,c_emergent,ros_remanent,wortis}. 
However, the general definitions of LIOMs may remain somewhat abstract and elusive for the newcomers in the field, so that it seems useful to have a more concrete picture 
in the simplest limit, namely in the strong disorder limit deep in the MBL-phase, where the LIOMs remain perturbatively close to the real-space degrees of freedom defining the model. 
In addition, the notion of LIOMs is often used at a qualitative level to explain the behavior of various observables, so it is important to discuss the quantitative link
in this strong disorder limit with other signatures of the MBL-phase. The goal of this paper is thus to revisit the Strong Disorder perturbative approach 
 for the Local Integrals of Motion from this perspective and to give explicit calculations up to second order for various quantum spin models.

The paper is organized as follows.
In section \ref{deflioms}, 
we recall how the pseudo-spins can be constructed from the true spins by the unitary transformation that diagonalizes the Hamiltonian.
In section \ref{expreal}, we describe how the expansion of the true spins in the Pauli basis of the pseudo-spins is related to the matrix elements of a single spin operator in the eigenstate basis and to the dynamics of the local magnetizations. 
In section \ref{explioms}, we mention the reciprocal expansion of the pseudo-spins in terms of the true spins. 
In section \ref{perturbation}, the lowest order of the strong disorder perturbative expansion for the pseudo-spins is described in detail.
The application to the random field XXZ chain is given in section \ref{twobody}.
The application to the toy model considered in \cite{c_toy,c_pseudocriti} is studied in section \ref{toy}
to analyze the stability of the MBL-phase.
Our conclusions are summarized in section \ref{conclusion}.
Finally, the Appendix \ref{appendice} describes the non-perturbative notion of LIOMs for the trivial case involving only two spins,
while the Appendix \ref{app_levy} contains some useful results on L\'evy sums of correlated variables.

\section{ Definition of Lioms in the strong disorder limit }

\label{deflioms}

\subsection{ Random quantum spin Models  }

Let us consider a model of $N=L^d $ quantum spins $\sigma_i$ 
described by 
 the hermitian Pauli matrices at each site
\begin{eqnarray}
\sigma^{(0)}  = Id =
 \begin{pmatrix}
1 & 0 \\
0 & 1 
\end{pmatrix}
\ \ \ \ \ 
\sigma^{x} 
= \begin{pmatrix}
0 & 1 \\
1 & 0 
\end{pmatrix}
\ \ \ \ \ 
\sigma^{y}  =
 \begin{pmatrix}
0 & -i \\
i & 0 
\end{pmatrix}
\ \ \ \ \ 
\sigma^{z} && =
 \begin{pmatrix}
1 & 0 \\
0 & -1 
\end{pmatrix}
\label{pauli}
\end{eqnarray}

with an Hilbert space of size
\begin{eqnarray}
{\cal N} = 2^N = 2^{L^d}
\label{nhilbert}
\end{eqnarray}

The Hamiltonian  can be decomposed into a diagonal part and an off-diagonal part in the $\sigma^z$ basis
\begin{eqnarray}
H && =H^{diag}+H^{off}
\label{hmbl}
\end{eqnarray}
The diagonal part $H^{diag} $ contains disorder variables such as random fields $h_j$ that are drawn with some continuous distribution to avoid any exact degeneracy in the spectrum.

\subsection{ Unitary transformation diagonalizing the Hamiltonian }

When the off-diagonal part vanishes $H^{off}=0$, the ${\cal N} = 2^N$ eigenstates of $H^{diag}$ are simply labelled by the eigenvalues of the $\sigma^z_j$
\begin{eqnarray}
\vert \psi^{(0)}_{S_1,..,S_{N}} > && \equiv \vert \sigma_1^z=S_1 , \sigma_2^z=S_2 ,.... , \sigma_N^z=S_N >
\label{psizero}
\end{eqnarray}
where the random energies
\begin{eqnarray}
E^{(0)}_{S_1,..,S_{L}} =  <  S_1 S_2 .... S_N \vert H^{diag} \vert S_1 S_2 .... S_N >
\label{e0}
\end{eqnarray}
are non-degenerate as a consequence of the random fields.
Then each eigenstate can be followed via the non-degenerate perturbation theory in $H^{off}$,
and this defines a unitary transformation between the basis of unperturbed eigenstates and the basis of perturbed eigenstates
\begin{eqnarray}
&& \vert \psi_{S_1,..,S_{N}} > = U \vert S_1 S_2 .... S_N > = \sum_{S_1',..,S_N'}    \vert S_1' S_2' .... S_N' >< S_1' S_2' .... S_N'\vert  U \vert S_1 S_2 .... S_N >
\label{defU}
\end{eqnarray}
It is then interesting to consider the action of this unitary transformation of the spin operators with $a=x,y,z$ \cite{emergent_vidal,serbyn_criterion,ros_remanent}
\begin{eqnarray}
\tau_n^a && = U \sigma_n^{a} U^{\dagger} 
\label{deftau}
\end{eqnarray}
because these pseudo-spins $ \tau_n$ inherit the commutation relations of the true spins $\sigma_n$.

\subsection{ Interpretation of the eigenstates in terms of the pseudo-spins}

The unperturbed eigenstate of Eq. \ref{psizero} is associated to the projector 
\begin{eqnarray}
 \vert S_1 S_2 .... S_N > < S_1 S_2 .... S_N \vert = \prod_{i=1}^N \left( \frac{1+ S_i \sigma_i^z}{2} \right)
\label{psizeroproj}
\end{eqnarray}
Its transformation via the unitary transformation $U$ reads using the definition of pseudo-spins in Eq \ref{deftau}
\begin{eqnarray}
U \vert S_1 S_2 .... S_N > < S_1 S_2 .... S_N \vert U^{\dagger}= \prod_{i=1}^N \left( \frac{1+ S_i \tau_i^z}{2} \right)
=  \vert \tau_1^z=S_1, ... , \tau_N^z=S_N >  <\tau_1^z=S_1, .... , \tau_N^z=S_N \vert
\label{uprojtau}
\end{eqnarray}
and thus corresponds to the projector on the state  $\tau_i^z=S_i$. 
On the other hand, by definition of  the unitary transformation of Eq. \ref{defU}, 
this coincides with the projector onto the perturbed eigenstate $ \vert \psi_{S_1,..,S_{N}} >$
\begin{eqnarray}
U \vert S_1 S_2 .... S_N > < S_1 S_2 .... S_N \vert U^{\dagger}= \vert \psi_{S_1,..,S_{N}} > < \psi_{S_1,..,S_{N}}  \vert
\label{uprojpsi}
\end{eqnarray}
By identification, one obtains that the perturbed eigenstates
 $\vert \psi_{S_1,..,S_{N}} > $ corresponds to the eigenstate of the pseudo-spin operators $\tau_i^z=S_i$
\begin{eqnarray}
 \vert \psi_{S_1,..,S_{N}} > =  \vert \tau_1^z=S_1, \tau_2^z= S_2, .... , \tau_N^z=S_N >
\label{eigenlabelrtau}
\end{eqnarray}
The $N$ pseudo-spins $\tau^z_j$ are thus a very convenient way to label the $2^N$ eigenstates.

\subsection{ Hamiltonian in terms of the pseudo-spins}

The Hamiltonian is diagonal in the basis of the eigenstates labelled by the pseudo-spins (Eq. \ref{eigenzeropertau})
\begin{eqnarray}
 H = \sum_{T_1=\pm,...,T_N=\pm } E_{T_1,..,T_N}  \vert \tau_1^z=T_1, \tau_2^z= T_2, .... , \tau_N^z=T_N >
< \tau_1^z=T_1, \tau_2^z= T_2, .... , \tau_N^z=T_N \vert
\label{eigenzeropertau}
\end{eqnarray}
and can be thus rewritten at the operator level as
\begin{eqnarray}
H =  \sum_{a_1=0,z} ...  \sum_{a_N=0,z}
{\cal H}_{a_1...a_N }  \tau_1^{(a_1)}  \tau_2^{(a_2)}  ...  \tau_N^{(a_N)} 
\label{hldiag}
\end{eqnarray}
where the ${\cal N}=2^N$ coefficients ${\cal H}_{a_1...a_N } $ can be computed to reproduce the $2^N$ energies $E_{T_1,..,T_N} $ 

\subsection{ Pauli basis of the pseudo-spins}

More generally, the notion of the basis of Pauli matrices for operators is very useful \cite{emergent_vidal,keating,c_toda}.
Any operator $O$ can be thus expanded 
in the basis of Pauli matrices of the $N$ pseudo-spins as \cite{emergent_vidal}
\begin{eqnarray}
O =  \sum_{a_1=0,x,y,z} ...  \sum_{a_N=0,x,y,z} 
{\cal O}(a_1...a_N ) \tau_1^{(a_1)}  \tau_2^{(a_2)}  ...  \tau_N^{(a_N)} 
\equiv  \sum_{\vec a} {\cal O}(\vec a )  \tau_1^{(a_1)}  \tau_2^{(a_2)}  ...  \tau_N^{(a_N)} 
\label{otau}
\end{eqnarray}
where the $4^N$ coefficients labelled by $\vec a=(a_1,..,a_N)$ read
\begin{eqnarray}
{\cal O}(\vec a )  = \frac{1}{2^N} Tr ( O \tau_1^{(a_1)}  \tau_2^{(a_2)}  ...  \tau_N^{(a_N)}  ) 
\label{otaua}
\end{eqnarray}

\section{Expansion of the real spins in terms of the pseudo-spins }

\label{expreal}

It is interesting to consider the expansion of Eq. \ref{otau}
for the real-spin-operator ${\cal O}=\sigma_n^z$
\begin{eqnarray}
\sigma_n^z =  \sum_{\vec a} {\cal S}_{n} (\vec a )  \tau_1^{(a_1)}  \tau_2^{(a_2)}  ...  \tau_N^{(a_N)} 
\label{sigmaexptau}
\end{eqnarray}

\subsection{ Properties of the coefficients }

The  coefficients
\begin{eqnarray}
{\cal S}_n( \vec a )  = \frac{1}{2^N} Tr ( \sigma_n^z  \tau_1^{(a_1)}  \tau_2^{(a_2)}  ...  \tau_N^{(a_N)}    ) 
\label{sna}
\end{eqnarray}
are real
\begin{eqnarray}
({\cal S}_n^*(\vec a ) )^* = {\cal S}_n(\vec a )
\label{nnareal}
\end{eqnarray}
as a consequence of hermiticity of Pauli matrices, while the zero-coefficient vanishes as a consequence of the vanishing trace of Pauli matrices 
\begin{eqnarray}
{\cal S}_n(\vec 0 )  = \frac{1}{2^N} Tr ( \sigma_n^z    ) =0
\label{snazero}
\end{eqnarray}

The condition of identity for the square
\begin{eqnarray}
I = (\sigma_n^z)^2 =  \sum_{\vec a}  {\cal S}_n (\vec a )  \sum_{\vec a'}  {\cal S}_n^b (\vec a ' ) 
\tau_1^{(a_1)} \tau_1^{(a_1')}  \tau_2^{(a_2)} \tau_2^{(a_2')}  ...  \tau_N^{(a_N)} \tau_N^{(a_N')} 
\label{idsquare}
\end{eqnarray}
yields in particular by taking the trace 
\begin{eqnarray}
1 =\frac{1}{2^N} Tr (  (\sigma_n^z)^2) =  \sum_{\vec a}  {\cal S}^2_n (\vec a ) 
\label{tracenorma}
\end{eqnarray}
that the sum of the square of the $4^N$ coefficients ${\cal S}_n (\vec a )  $ is normalized to unity.

\subsection{ Matrix elements of a single real spin operator in the eigenstate basis }

\label{matrixelsigma}

The behavior of matrix elements of local operators in the eigenstate basis
has been proposed as an important criterion to distinguish 
the Many-Body-Localized phase from the Delocalized phase \cite{serbyn_criterion,serbyn_dyson,c_dysonBM,c_mblstrongmultif}.
It is thus interesting to consider the consequence of the expansion Eq. \ref{sigmaexptau}
on the matrix element of the single real-spin operator $\sigma_n^z$ 
between two eigenstates labelled as $\vert \vec T> = \vert \tau_1^z=T_1,..,\tau_N^z=T_N>$ and $\vert \vec T ' > = \vert \tau_1^z=T_1' ,..,\tau_N^z=T_N' >$
\begin{eqnarray}
 < \vec T \vert \sigma_n^z \vert \vec T '> && =  <  T_1,..,T_N  \vert \sum_{\vec a} {\cal S}_{n} (\vec a )  \tau_1^{(a_1)}  \tau_2^{(a_2)}  ...  \tau_N^{(a_N)}  \vert  T_1' ,..,T_N' > 
 = \sum_{\vec a} {\cal S}_{n} (\vec a ) \prod_{i=1}^N < T_i \vert \tau_i^{(a_i)} \vert T_i' >
\label{mni}
\end{eqnarray}

 To characterize the amplitude of these matrix elements, 
it is convenient to introduce the Edwards-Anderson matrix \cite{c_dysonBM,c_mblstrongmultif}
\begin{eqnarray}
 Q_{\vec T,\vec T'} \equiv   \vert < \vec T \vert \sigma_n^z \vert   \vec T '>  \vert^2=
< \vec T \vert \sigma_n^z \vert  \vec T '>< \vec T ' \vert \sigma_n^z \vert \vec T>
\label{qttprime}
\end{eqnarray}
that has the nice property to be doubly stochastic,
 i.e. it is a square matrix of size ${\cal N} \times {\cal N}$
of non-negative real numbers,
where the sums over any row or any column is unity
\begin{eqnarray}
\sum_{\vec T}  Q_{\vec T,\vec T'}   = 1 =\sum_{\vec T'}  Q_{\vec T,\vec T'} 
\label{bistochloci}
\end{eqnarray}
as a consequence of the completeness identity for the eigenstate basis
and the Pauli matrix identity $(\sigma_n^z)^2=I $.

The normalization of Eq. \ref{bistochloci} means that for a fixed eigenstate $\vert \vec T>$,
the ${\cal N}$ numbers $Q_{\vec T,\vec T'} $
can be interpreted as ${\cal N} $ weights normalized to unity.
Their statistics can be characterized by the multifractal formalism as follows (see more details in \cite{c_mblstrongmultif}) :
the number of weights of order $Q_{\vec T,\vec T'} \propto {\cal N}^{-\alpha} $ among the ${\cal N} $ weights scales as
\begin{eqnarray}
{\rm Number } (Q_{\vec T,\vec T'} \propto {\cal N}^{-\alpha} ) \propto {\cal N}^{f(\alpha)} 
\label{multif}
\end{eqnarray}
Equivalently, the generalized moments can be computed via a saddle-point analysis in the exponent $\alpha$
\begin{eqnarray}
Y_q \equiv \sum_{\vec T' } \vert < \vec T  \vert \sigma_{n}^z \vert \vec T '> \vert^{2q}
&& = \int d\alpha {\cal N}^{f(\alpha)-q \alpha} = {\cal N}^{-\tau(q) }
\label{yqdef}
\end{eqnarray}
with the usual Legendre transformation between the multifractal spectrum $f(\alpha)$ and the exponents $\tau(q)$
\begin{eqnarray}
f(\alpha)-q \alpha && = -\tau(q) 
\nonumber \\
f'(\alpha) -q && =0
\label{legendre}
\end{eqnarray}

When the state $\vert \vec T >$ is in the middle of the spectrum, the 
the Ergodic phase where the Eigenstate Thermalization Hypothesis (E.T.H.) \cite{deutsch,srednicki,nature,mite,rigol} holds
is characterized by the monofractal
\begin{eqnarray}
f^{ETH}(\alpha)=\delta(\alpha-1)
\label{feth}
\end{eqnarray}
i.e. there is an extensive number $O({\cal N})$ of weights that are of order $Q_{\vec T,\vec T'} \propto 1/{\cal N} $,
so that the generalized moments scale linearly in $q$ as
\begin{eqnarray}
Y_q^{ETH}  \propto {\cal N}^{1-q }
\label{yqeth}
\end{eqnarray}
On the contrary in the limit where $H_{off}=0$ where the pseudo spins $\tau^z$ coincides with the true spins $\sigma^z$,
only one weight is non-zero $Q_{\vec T,\vec T'} =\delta_{\vec T, \vec T'}  $. 
More generally in the Many-Body-Localized phase, this weight remains finite
\begin{eqnarray}
Q^{MBL}_{\vec T,\vec T} =O(1)
\label{qttfinite}
\end{eqnarray}
i.e. the multifractal spectrum begins at $\alpha=0$ with $f^{MBL}(\alpha=0)=0$.
Introducing the linear slope around the origin
\begin{eqnarray}
f^{MBL}(\alpha)=q_c \alpha +o(\alpha)
\label{fmbl}
\end{eqnarray}
one obtains that
the saddle-point calculation of Eq \ref{yqdef} is dominated by this boundary $\alpha=0$ 
for $q>q_c$ yielding finite generalized moments
\begin{eqnarray}
Y_{q>q_c}^{MBL} = O(1)
\label{yqmbl}
\end{eqnarray}

\subsection{ Dynamics of the local magnetizations  }

\label{dynmagneti}

The dynamics of the local magnetizations
\begin{eqnarray}
<\sigma_n^z>_t = Tr (\sigma_n^z  e^{-i Ht } \rho(t=0)e^{i Ht } )
\label{dynmagn}
\end{eqnarray}
can be considered as the most important criterion to distinguish 
the Many-Body-Localized phase from the Delocalized phase.
To make the link with the above framework, it is convenient to focus on the simple initial density matrix \cite{emergent_vidal}
\begin{eqnarray}
\rho(t=0)= \frac{1+\sigma^z_{n_0}}{2^N} 
\label{rho0sigman_0}
\end{eqnarray}
corresponding to magnetization unity on the site $n_0$ and zero magnetization on all the other sites
\begin{eqnarray}
<\sigma_n^z>_{t=0} = Tr( \sigma_n^z \frac{1+\sigma^z_{n_0}}{2^N} ) =\delta_{nn_0}
\label{mini}
\end{eqnarray}

The expansion upon the basis of eigenstates $\vert \vec T > =\vert T_1,..,T_N>$
\begin{eqnarray}
<\sigma_n^z>_t = \sum_{\vec T}  \sum_{\vec T' } < \vec T \vert  \sigma_n^z \vert \vec T' > < \vec T ' \vert \rho(0)\vert \vec T >
e^{i (E_{\vec T}   -E_{\vec T'}) t }
\label{iphase}
\end{eqnarray}
yields that the time-average eliminates the off-diagonal terms $\vec T ' \ne \vec T$ 
\begin{eqnarray}
 \frac{1}{t_{max}} \int_0^{t_{max}} dt <\sigma_n^z>_t \opsimeq_{t_{max} \to +\infty}
 \sum_{\vec T}  < \vec T \vert  \sigma_n^z \vert \vec T > < \vec T  \vert \rho(0)\vert \vec T >
  \equiv m_n^{\infty}
\label{timeav}
\end{eqnarray}
For the initial condition of Eq \ref{rho0sigman_0}, one obtains that these magnetization $m_n^{\infty} $
are directly related to the matrix elements discussed above
\begin{eqnarray}
 m_n^{\infty} =
 \frac{1}{2^N}  \sum_{\vec T}  < \vec T \vert  \sigma_n^z \vert \vec T >  < \vec T  \vert \sigma^z_{n_0} \vert \vec T >
\label{mninfty}
\end{eqnarray}
In particular, an important property of the Many-Body-Localized phase
is the presence of some memory of the initial magnetization unity on the site $n=n_0$
\begin{eqnarray}
 m_{n=n_0}^{\infty} = \frac{1}{2^N}  \sum_{\vec T}  (< \vec T \vert  \sigma_n^z \vert \vec T >)^2 =  \frac{1}{2^N}  \sum_{\vec T} Q_{\vec T, \vec T}
\label{mpinfty}
\end{eqnarray}
which involves an average over the eigenstate $\vec T $ of the diagonal terms $Q_{\vec T, \vec T} $ of Eq. \ref{qttprime}.

\section{ Expansion of the LIOMs in terms of the real spins}

\label{explioms}

\subsection{Expansion of the pseudo-spins in the Pauli basis of the real spins }

Reciprocally, it is interesting to consider the expansion of the LIOMs $\tau_n^z$ in the Pauli basis of the true spins $\sigma_i$
\begin{eqnarray}
\tau_n^z =  \sum_{\vec a} {\cal T}_n(\vec a )  \sigma_1^{(a_1)}  \sigma_2^{(a_2)}  ...  \sigma_N^{(a_N)} 
\label{tauexpsigma}
\end{eqnarray}
where the  coefficients
\begin{eqnarray}
{\cal T}_n( \vec a )  = \frac{1}{2^N} Tr ( \tau_n^z  \sigma_1^{(a_1)}  \sigma_2^{(a_2)}  ...  \sigma_N^{(a_N)}   ) 
\label{tna}
\end{eqnarray}
are real $ {\cal T}_n^*(\vec a ) = {\cal T}_n(\vec a )$, satisfy ${\cal T}(\vec 0 )  =0 $ 
and the normalization similar to Eq. \ref{tracenorma}
\begin{eqnarray}
1 =\frac{1}{2^N} Tr (\tau_n^z)^2 =  \sum_{\vec a}  {\cal T}^2_n(\vec a )  
\label{normata}
\end{eqnarray}

\subsection{Overlap between one pseudo-spin and the corresponding real spin }

The overlap $ O_n$ between the real spin $\sigma_n^z$ and the corresponding pseudo-spin $\tau_n^z=U \sigma_n U^{\dagger} $ (Eq. \ref{deftau})
appears as the coefficient associated to $(a_n=z; a_{i \ne n}=0 )$ either in the expansion of Eq. \ref{sigmaexptau}
or in the expansion of Eq. \ref{tauexpsigma}
\begin{eqnarray}
O_n \equiv \frac{1}{2^N} Tr ( \tau_n^z  \sigma_n^z  ) =  {\cal S}_n( a_n=z; a_{i \ne n}=0  )= {\cal T}_n( a_n=z; a_{i \ne n}=0  )  
\label{overlan_0}
\end{eqnarray}
It is the direct measure of the locality of the pseudo-spin $\tau_n^z$ : it remains finite in the Many-Body-Localized phase
\begin{eqnarray}
O_n^{MBL} = O(1)
\label{overlapmbl}
\end{eqnarray}
while it vanishes in the thermodynamic limit in the delocalized phase.

\subsection{ Dynamics from a fixed initial condition }

\label{liomsconserved}

The $N$ operators $\tau_n^z$ commuting with each other and with the Hamiltonian
represent $N$ elementary integrals of motion, from which one can generate all the other ones by linear combination of products $\tau_n^z \tau_m^z ...$
(up to the $2^N$ projectors on the eigenstates of Eq. \ref{uprojtau}). Within the strong disorder perturbative expansion, the $N$ pseudospins $ \tau_n^z$ 
are thus clearly the extensive set of the most 'local' Integrals of Motion that one can construct.

If one starts from a given initial condition in the physical $\sigma^z$ basis $\vert \psi(t=0)>=\vert S_1.,,S_N>$ \cite{ros_remanent},
the values of these integrals of motion read
\begin{eqnarray}
<\tau^z_n >_t = <\tau^z_n >_{t=0} =  < S_1,..,S_N \vert  \tau^z_n \vert S_1,..,S_N>
\label{1config}
\end{eqnarray}

\section{Self-consistent first-order perturbative expansion  }

\label{perturbation}

\subsection{ Standard first order perturbation theory   }

For $H^{off}=0$, the ${\cal N}=2^{N}$ eigenstates of $H^{diag}$ are simply given by the tensor products of Eq. \ref{psizero}.
The usual non-degenerate first order perturbation theory yields the eigenstates using the simplified notation $\vert \vec S>=\vert S_1,..,S_N>$
\begin{eqnarray}
&& \vert \psi^{(0+1)}_{\vec S} > 
=  \vert \vec S >  + \sum_{ \vec S' \ne \vec S } \vert\vec S' > 
 \frac{ < \vec S' \vert H^{off}\vert \vec S > }{ E^{(0)}_{\vec S} -  E^{(0)}_{\vec S '}}
\label{eigenzeroper}
\end{eqnarray}

This result can be directly translated for the perturbative expansion of the unitary transformation $U$ 
\begin{eqnarray}
U= 1+ U_1
\label{uper}
\end{eqnarray}
describing this change of basis (Eq. \ref{defU})
\begin{eqnarray}
&& \vert \psi^{(0+1)}_{\vec S } > 
= (  1+ U_1)  \vert \vec S > 
\label{eigenzeroperu}
\end{eqnarray}
The identification with Eq. \ref{eigenzeroper}
yields the matrix elements of the first order $U_1$
\begin{eqnarray}
 < \vec S ' \vert U_1 \vert \vec S> = \frac{ < \vec S' \vert H^{off}\vert \vec S > }{ E^{(0)}_{\vec S} -  E^{(0)}_{\vec S '}}
\label{u1res}
\end{eqnarray}
At the operator level, this can be rewritten as the usual commutator equation for $U_1$
\begin{eqnarray}
 [ U_1, H^{diag} ] = H^{off}
\label{commueta1}
\end{eqnarray}

The transformation of any operator $O$ then reads at this order
\begin{eqnarray}
{\tilde O} && = U O U^{\dagger} =  (1+U_1+..) O (1-U_1+..) 
=  O + [U_1,O ]  +...
\label{Otrans}
\end{eqnarray}
In particular, the pseudo-spins $\tau_n^z=U \sigma_n^{z} U^{\dagger} $ (Eq. \ref{deftau}) display the following perturbative expansion around the real spins
\begin{eqnarray}
\tau_{n(0+1)}^z  = \sigma_n^z + [ U_1, \sigma_n^z] 
\label{tauper012}
\end{eqnarray}
Reciprocally, the real spins $\sigma_n^z=U^{\dagger} \sigma_n^{z} U $ can be expanded around the pseudo-spins
\begin{eqnarray}
\sigma_{n(0+1)}^z  = \tau_n^z - [ U_1, \tau_n^z] 
\label{sigmaper012}
\end{eqnarray}

\subsection{ Self-consistent first order perturbation theory for the eigenstates  } 

In random systems, resonances may appear in the first-order eigenstate of Eq. \ref{eigenzeroper}
when the amplitude $\vert \frac{ < \vec S' \vert H^{off}\vert \vec S > }{ E^{(0)}_{\vec S} -  E^{(0)}_{\vec S '}} \vert$ is not small.
The simplest way to take into account these rare possible resonances is to introduce the normalized version of Eq. \ref{eigenzeroper}
\begin{eqnarray}
\vert \psi^{(0+1)norm}_{\vec S} >&&  \equiv \frac{ \vert \psi^{(0+1)}_{\vec S} > }{\sqrt{ < \psi^{(0+1)}_{\vec S} \vert \psi^{(0+1)}_{\vec S} > } } 
= \frac{   \vert \vec S >  + \displaystyle \sum_{ \vec S' \ne \vec S } \vert\vec S' > 
 \frac{ < \vec S' \vert H^{off}\vert \vec S > }{ E^{(0)}_{\vec S} -  E^{(0)}_{\vec S '}}  }
{\sqrt{ 1+ \displaystyle \sum_{ \vec S' \ne \vec S } \left \vert
 \frac{ < \vec S' \vert H^{off}\vert \vec S > }{ E^{(0)}_{\vec S} -  E^{(0)}_{\vec S '}} \right\vert^2 }  } 
\label{eigenzeropernorm}
\end{eqnarray}
as explained in detail in \cite{us_stronglevy,c_toy} for Anderson-Localization and Many-Body-Localization respectively.
What happens at the technical level is that the term $\vert \frac{ < \vec S' \vert H^{off}\vert \vec S > }{ E^{(0)}_{\vec S} -  E^{(0)}_{\vec S '}} \vert^2$
of the denominator, although formally of second order, is actually of first order in the off-diagonal couplings, as a consequence of resonances
\cite{us_stronglevy,c_toy}. Eq. \ref{eigenzeropernorm} can be thus considered as the self-consistent first order perturbative expression for the eigenstate.

\subsection{ Self-consistent first order perturbation theory for the LIOMs and for the spins operators } 

Here we wish to apply the same idea to LIOMS that are operators, so that the appropriate norm is the Froebenius or Hilbert-Schmidt norm,
based on the inner product of two operators
\begin{eqnarray}
(A,B) = \frac{1}{2^N}  Tr(A^{\dagger} B)
\label{inner}
\end{eqnarray}

So the normalized version of Eq. \ref{tauper012} reads
\begin{eqnarray}
\tau_{n(0+1)norm}^z  \equiv \frac{ \tau_{n(0+1)}^z } { \sqrt{ ( \tau_{n(0+1)}^z,\tau_{n(0+1)}^z ) }  } 
 = \frac{   \sigma_n^z + [ U_1, \sigma_n^z] } { \sqrt{ 1+ \frac{1}{2^N} Tr ([ U_1, \sigma_n^z]^2 )   }  }
\label{tauper01norm}
\end{eqnarray}

Similarly, the normalized version of Eq. \ref{sigmaper012}  reads
\begin{eqnarray}
\sigma_{n(0+1)norm}^z  \equiv \frac{ \sigma_{n(0+1)}^z } { \sqrt{ ( \sigma_{n(0+1)}^z,\sigma_{n(0+1)}^z ) }  } 
 = \frac{   \tau_n^z - [ U_1, \tau_n^z] } { \sqrt{ 1+ \frac{1}{2^N} Tr ([ U_1, \tau_n^z]^2 )   }  }
= \frac{   \tau_n^z - [ U_1, \tau_n^z] } { \sqrt{ 1+ \frac{1}{2^N} \displaystyle \sum_{\vec T} < \vec T \vert [ U_1, \tau_n^z]^2 \vert \vec T >   }  }
\label{sigmaper01norm}
\end{eqnarray}
so that the memory of the initial magnetization of Eq. \ref{mpinfty} becomes 
\begin{eqnarray}
 m_{n=n_0}^{\infty} = \frac{1}{2^N}  \sum_{\vec T}  (< \vec T \vert \sigma_{n(0+1)}^z  \vert \vec T >)^2  
= \frac{  1 } {  1+ \frac{1}{2^N} Tr ([ U_1, \tau_n^z]^2 )    }
= \frac{  1 } {  1+ \frac{1}{2^N} \displaystyle \sum_{\vec T} < \vec T \vert [ U_1, \tau_n^z]^2 \vert \vec T >     }
\label{mpinftynorm}
\end{eqnarray}

In the two following sections, this self-consistent first order perturbation theory is applied to specific models.

\section { Application to the random field XXZ chain }

\label{twobody}

In this section, we consider the nearest-neighbor XXZ chain with random fields $h_j$, corresponding to the diagonal part
\begin{eqnarray}
H^{diag} && = \sum_j h_j \sigma_j^z  +\sum_{j} \Delta \sigma_j^z \sigma_{j+1}^z 
\label{diag2body}
\end{eqnarray}
and the off-diagonal part
\begin{eqnarray}
H^{off} && =  \sum_{j}  J (\sigma_j^x \sigma_{j+1}^x + \sigma_j^y \sigma_{j+1}^y  )
= \sum_{j}   2J (\sigma_j^+ \sigma_{j+1}^- + \sigma_j^- \sigma_{j+1}^+  )
\label{hoff2body}
\end{eqnarray}
The isotropic case $J=1=\Delta$ is the most studied Many-Body-Localization model where many numerical results are available
\cite{kjall,alet,alet_dyn,luitz_tail,badarson_signa,auerbach,znidaric_dephasing,prelo_dyn,znidaric_lindblad,luitz_bimodal,garcia,luitz_anomalous,luitz_operator,luitz_information}.

\subsection{ Perturbation in the off-diagonal coupling $J$  }

When applied to the XXZ Hamiltonian of Eqs \ref{diag2body} , \ref{hoff2body},
the first order perturbation theory described in the previous section
yields the following first order contribution $U_1$ for the unitary transformation $U=1+U_1$
\begin{eqnarray}
U_1 = \sum_{j_0} 
 \theta_{j_0,j_0+1}(\tau_{j_0-1}^z ; \tau_{j_0+2}^z ) 
(\tau_{j_0}^+ \tau_{j_0+1}^- -  \tau_{j_0}^- \tau_{j_0+1}^+ )
\label{eta1sigma}
\end{eqnarray}
with the notation
\begin{eqnarray}
 \theta_{j_0,j_0+1}(\tau_{j_0-1}^z ; \tau_{j_0+2}^z ) && \equiv
  \frac{J }{\left( h_{j_0+1}+  \Delta \tau_{j_0+2}^z \right)  -  \left(h_{j_0}+  \Delta \tau_{j_0-1}^z \right)   }
\label{thetaphi}
\end{eqnarray}

The commutator appearing in Eq. \ref{sigmaper01norm} reads
\begin{eqnarray}
 [ U_1, \tau_n^z ]
&& = 2 \theta_{n-1,n}(\tau_{n-2}^z ; \tau_{n+1}^z ) ( \tau_{n_1}^+ \tau_{n}^- + \tau_{n_1}^- \tau_{n}^+ )
  -  2 \theta_{n,n+1}( \tau_{n-1}^z ; \tau_{n+2}^z) ( \tau_{n}^+ \tau_{n+1}^- +  \tau_{n}^- \tau_{n+1}^+ ) 
\label{sigman1thetaphi}
\end{eqnarray}
The diagonal part of its square
\begin{eqnarray}
Diag( [ U_1, \tau_n^z ]^2 )  
&& = 2 \theta^2_{n-1,n}(\tau_{n-2}^z ; \tau_{n+1}^z  )  (1- \tau_{n-1}^z \tau_{n}^z  )
+ 2   \theta^2_{n,n+1}( \tau_{n-1}^z ; \tau_{n+2}^z ) (1-\tau_{n}^z \tau_{n+1}^z )
\label{normataujz1er}
\end{eqnarray}
yields that the trace needed for the magnetization memory of Eq. \ref{mpinftynorm} simplifies into the following sum
\begin{eqnarray}
  {\cal T}  &&  \equiv  \frac{1}{2^N} Tr ([ U_1, \tau_n^z]^2 )  = \frac{1}{2^N}  \sum_{\vec T} < \vec T \vert [ U_1, \tau_n^z]^2 \vert \vec T >     
=  \frac{2}{2^N}   \sum_{T_1,..,T_N}   ( \theta^2_{n-1,n}(\tau_{n-2}^z ; \tau_{n+1}^z  )  + \theta^2_{n,n+1}( \tau_{n-1}^z ; \tau_{n+2}^z )  ) 
\nonumber \\
&& =  \frac{1}{2}   \sum_{T_{n-2}=\pm1, T_{n+1}=\pm 1}   \theta^2_{n-1,n}(T_{n-2} ; T_{n+1}  )  
+  \frac{1}{2}   \sum_{T_{n-1}=\pm1, T_{n+2}=\pm 1}   \theta^2_{n,n+1}( T_{n-1} ; T_{n+2} )
\nonumber \\
&& =  \frac{J^2}{2} \left(
 \frac{2 }{\left( h_{n}-h_{n-1}  \right)^2 } 
+ \frac{1 }{\left( h_{n}-h_{n-1} + 2 \Delta   \right)^2 } 
+  \frac{1 }{\left( h_{n}-h_{n-1} -2  \Delta   \right)^2 } 
\right)
 \nonumber \\
 &&
+  \frac{J^2}{2} \left(  
 \frac{2 }{\left( h_{n+1}-h_n  \right)^2 } 
+  \frac{1 }{\left( h_{n+1}-h_n +  2 \Delta  \right)^2 } 
+  \frac{1 }{\left( h_{n+1}-h_n - 2  \Delta   \right)^2 } 
 \right)
\label{sigma2}
\end{eqnarray}
that involves only the three consecutive random fields $(h_{n-1},h_n,h_{n+1})$.

\subsection{Magnetization  memory as lowest order in the off-diagonal coupling $J$ }

The magnetization memory of Eq. \ref{mpinftynorm} is directly related to the positive sum ${\cal T}$ introduced in Eq. \ref{sigma2}
\begin{eqnarray}
 m_{n=n_0}^{\infty} = \frac{  1 } {  1+  {\cal T}  } 
\label{mpinftynorms}
\end{eqnarray}
As detailed in Appendix \ref{app_levy}, its disorder-averaged value is not of order $J^2$ 
but of order $\vert J \vert$ as a consequence of the L\'evy distribution of ${\cal T}$ induced by the resonances (Eq \ref{mav})
\begin{eqnarray}
\overline{ m_{n=n_0}^{\infty} } 
&& = 1 -   \vert J \vert {\cal A} + o( \vert J_i \vert ) 
\label{mavxxz}
\end{eqnarray}
where the amplitude ${\cal A}$ can be computed (Eq \ref{mav}) from the properties of the individual terms appearing in Eq. \ref{sigma2}.
This behavior in $\vert J \vert$ was already found in \cite{ros_remanent} and can be also computed axactly (i.e. without perturbation theory)
in the trivial two-spins example described in the Appendix \ref{appendice} without perturbation theory (Eqs \ref{mpinfty2sres} \ref{mpinfty2sresper} \ref{mpinfty2sresperav}).

\section{ Application to the toy model \cite{c_toy,c_pseudocriti}}

\label{toy}

In this section, we consider the toy model \cite{c_toy,c_pseudocriti} where the diagonal part 
\begin{eqnarray}
 H^{diag} && = \sum_{i=1}^N h_i \sigma_i^z
\label{toydiag}
\end{eqnarray}
contains only random fields that are distributed with the Gaussian distribution of variance $W^2$ 
\begin{eqnarray}
P(h)= \frac{1}{\sqrt{ 2 \pi W^2}} e^{- \frac{h^2}{2 W^2}}
\label{gauss}
\end{eqnarray}
while the off-diagonal part contains hopping to all other configurations of the Hilbert space
\begin{eqnarray}
H^{off} && = \sum_{k=1}^N \sum_{1 \leq i_1<i_2<..<i_k \leq N} J_{i_1,..,i_k} \sigma_{i_1}^x \sigma_{i_2}^x .. \sigma_{i_k}^x 
\label{toyoff}
\end{eqnarray}
The question is how weak the behavior of the off-diagonal couplings $J_{i_1,..,i_k} $ should be 
in order to allow the existence of the Many-Body-Localized phase.
In \cite{c_toy}, the analysis was based on the entanglement entropy for the one-dimensional case.
Here we wish instead to analyze this type of model in dimension $d$ from the point of view of the pseudo-spins.

\subsection{ Perturbation in the off-diagonal couplings   }

The unitary transformation $U=1+U_1$ reads at first order in the off-diagonal couplings
\begin{eqnarray}
 U_1 && = \sum_{k=1}^N \sum_{1 \leq i_1<i_2<..<i_k \leq N} \frac{ J_{i_1,..,i_k} }{-2 \sum_{p=1}^k h_{i_p} \sigma_{i_p}^z } \sigma_{i_1}^x \sigma_{i_2}^x .. \sigma_{i_k}^x 
\label{etatoy}
\end{eqnarray}
To simplify the notations, let us focus on the first spin $\sigma_{n=1}^z$ and compute the 
commutator of Eq. \ref{sigmaper01norm}
\begin{eqnarray}
 [ U_1 ,\tau_{n=1}^z] && = \sum_{k=1}^N \sum_{1 = i_1<i_2<..<i_k \leq N} \frac{ J_{i_1,..,i_k} }{ \sum_{p=1}^k h_{i_p} \tau_{i_p}^z }  \tau_{1}^z \tau_{1}^x  \tau_{i_2}^x .. \tau_{i_k}^x 
\label{sigma1toy}
\end{eqnarray}
The diagonal part of its square
\begin{eqnarray}
Diag( [ U_1 ,\tau_{n=1}^z]^2)   && = \sum_{k=1}^N \sum_{1 = i_1<i_2<..<i_k \leq N} \left( \frac{ J_{i_1,..,i_k} }{ \sum_{p=1}^k h_{i_p} \tau_{i_p}^z }  \right)^2
\label{tau2toy}
\end{eqnarray}
yields that the trace needed for the magnetization memory of Eq. \ref{mpinftynorm} simplifies into the following sum
\begin{eqnarray}
 {\cal T}  &&  \equiv  \frac{1}{2^N} Tr ([ U_1, \tau_1^z]^2 )  = \frac{1}{2^N}  \sum_{\vec T} < \vec T \vert [ U_1, \tau_1^z]^2 \vert \vec T >     
=  \frac{1}{2^N}   \sum_{T_1,..,T_N}  \sum_{k=1}^N \sum_{1 = i_1<i_2<..<i_k \leq N} \left( \frac{ J_{i_1,..,i_k} }{ \sum_{p=1}^k h_{i_p} T_{i_p} } 
\right)^2
\nonumber
 \\
&& = \sum_{k=1}^N \sum_{1 = i_1<i_2<..<i_k \leq N}  \frac{1}{2^k}  \sum_{ T_1=\pm 1, T_{i_2}=\pm 1,..,T_{i_k=\pm 1} } \left( \frac{ J_{i_1,..,i_k} }{ \sum_{p=1}^k h_{i_p} T_{i_p} }  \right)^2
\label{trcarre}
\end{eqnarray}

\subsection{Memory of the local magnetization  }

The disorder-averaged value of the magnetization memory of Eq. \ref{mpinftynorm} reads in terms of Eq. \ref{trcarre}
\begin{eqnarray}
\overline{ m_{n=n_0=1}^{\infty} }
&& = \int_{-\infty}^{+\infty} dh_1 P(h_1)...  \int_{-\infty}^{+\infty} dh_N P(h_N) \frac{1}{1+ \displaystyle \sum_{k=1}^N \sum_{1 = i_1<..<i_k \leq N}  \frac{1}{2^k}  \sum_{ T_1=\pm 1, ..,T_{ik=\pm 1} } \left( \frac{ J_{i_1,..,i_k} }{ \sum_{p=1}^k h_{i_p} T_{i_p} }  \right)^2} 
\label{correcm1}
\end{eqnarray}
It is thus convenient to make the change of variables ${\tilde h_i}= h_{i} T_{i} $ and to use the Gaussian distribution of Eq. \ref{gauss}
to rewrite
\begin{eqnarray}
\overline{ m_{n=n_0=1}^{\infty} }
 && = \int_{-\infty}^{+\infty} d {\tilde h_1} P({\tilde h_1})...  \int_{-\infty}^{+\infty} d {\tilde h_N} P({\tilde h_N}) 
 \frac{1}{1+ \displaystyle \sum_{k=1}^N \sum_{1 = i_1<..<i_k \leq N}  \frac{1}{2^k}  \sum_{ T_1=\pm 1, ..,T_{ik=\pm 1} } \left( \frac{ J_{i_1,..,i_k} }{ \sum_{p=1}^k {\tilde h}_{i_p}  }  \right)^2} 
\nonumber \\
 && = \int_{-\infty}^{+\infty} d {\tilde h_1} P({\tilde h_1})...  \int_{-\infty}^{+\infty} d {\tilde h_N} P({\tilde h_N}) 
 \frac{1}{1+ \displaystyle \sum_{k=1}^N \sum_{1 = i_1<..<i_k \leq N}   \left( \frac{ J_{i_1,..,i_k} }{ \sum_{p=1}^k {\tilde h}_{i_p}  }  \right)^2} 
=\overline{ \frac{1}{1+ \Sigma } }
\label{correcm}
\end{eqnarray}
where
\begin{eqnarray}
\Sigma 
 && =   \sum_{k=1}^N \sum_{1 = i_1<..<i_k \leq N}   \left( \frac{ J_{i_1,..,i_k} }{ \sum_{p=1}^k {\tilde h}_{i_p}  }  \right)^2
\label{correcms}
\end{eqnarray}
is a L\'evy sum of correlated variables, whose statistical properties are discussed in Appendix \ref{app_levy}.
Since the sum of $k$ Gaussian variables (Eq \ref{gauss})
\begin{eqnarray}
E=\sum_{p=1}^k  {\tilde h}_{i_p}  
\label{esumk}
\end{eqnarray}
is distributed with the Gaussian distribution of variance $(k W^2)$
\begin{eqnarray}
\rho_{i_1,..,i_k}(E)= \frac{1}{\sqrt{ 2 \pi k W^2} } e^{- \frac{E^2}{2 k W^2}}
\label{gaussk}
\end{eqnarray}
the application of Eq. \ref{mav} to Eq. \ref{correcm} yields 
\begin{eqnarray}
\overline{ m_{n=n_0=1}^{\infty} }  
&&  = 1 - \pi \sum_{k=1}^N \sum_{1 = i_1<..<i_k \leq N}   \overline{  \vert  J_{i_1,..,i_k} \vert } \rho_{i_1,..,i_k}(0)  + o( \vert  J_{i_1,..,i_k} \vert ) 
\nonumber \\
&& = 1 - \frac{\sqrt  \pi}{ W \sqrt{ 2}   }  \sum_{k=1}^N k^{- \frac{1}{2}} \sum_{1 = i_1<..<i_k \leq N}  \overline{  \vert  J_{i_1,..,i_k} \vert  } + o( \vert  J_{i_1,..,i_k} \vert ) 
\label{mavtoy}
\end{eqnarray}

\subsection{ Criterion for the stability of the Many-Body-Localized phase }

\label{sec_physical}

The result of Eq. \ref{mavtoy} for the magnetization memory yields that
the criterion for the existence of the Many-Body-Localized phase is 
 the convergence in the thermodynamic limit $N \to +\infty$ 
of the following sum involving the averaged values $\overline{  \vert J_{i_1,..,i_k} \vert } $ 
\begin{eqnarray}
{\rm MBL\ \ phase \ \ criterion :  } \ \ \   \sum_{k=1}^N k^{- \frac{1}{2}} \sum_{1 = i_1<i_2<..<i_k \leq N}    \overline{  \vert J_{i_1,..,i_k} \vert } < +\infty
\label{yqmbltoy}
\end{eqnarray}

In \cite{c_toy,c_pseudocriti}, we have considered only the one-dimensional model
with the following specific form of the couplings $ J_{i_1,..,i_k} $ : they were assumed to depend only on the spatial range $r=(i_k-i_1)$ 
via some exponential decay governed by the control parameter $b$
and possibly some power-law prefactor governed by the parameter $a$.
\begin{eqnarray}
J_{i_1,..,i_k} = V \  \frac{2^{-b \vert i_k - i_1 \vert} v_{i_1,..,i_k}}{ \vert i_k - i_1 \vert^a} 
\label{gaussJ}
\end{eqnarray}
with $O(1)$ random variables $v_{i_1,..,i_k}$,
and the prefactor $V$ being the global vanishing amplitude to perform the perturbative expansion.
The properties of the entanglement entropy at the critical point $b_c=1$ could be then studied explicitly \cite{c_toy}.

Here it is thus interesting to discuss more generally the criterion of Eq. \ref{yqmbltoy} in dimension $d$ with $N=L^d$ spins.
It is clear that within the sum of Eq. \ref{yqmbltoy} containing a number of terms equal to the number of configurations 
\begin{eqnarray}
   \sum_{k=1}^N \sum_{1 = i_1<i_2<..<i_k \leq N}   =  \sum_{k=1}^N \binom {N} {k} = 2^N-1
\label{nconfig}
\end{eqnarray}
the most dangerous terms for the convergence are the terms corresponding to $k \simeq \frac{N}{2}$ 
that maximizes the binomial coefficient $\binom {N} {k} $ counting the number of choices for the $k$ positions $(i_1,..,i_k)$ within $N$
\begin{eqnarray}
    \binom {N} {\frac{N}{2} } \oppropto_{N \to +\infty}  2^N \sqrt{ \frac{2}{\pi N}}
\label{binomial}
\end{eqnarray}
that displays the same exponential behavior as the whole number of terms (Eq. \ref{nconfig}).
As a consequence, the important property of the couplings $J_{i_1,..,i_k}  $ for the MBL-transition
is the behavior for extensive resonances involving $k \simeq \frac{N}{2}$ spins $(i_1,..,i_{\frac{N}{2}})$ in a volume $N=L^d$.
Using the change of variable $k=\frac{N}{2} \left( 1+ \frac{x}{\sqrt N} \right)$ and the behavior of the binomial coefficient
\begin{eqnarray}
    \binom {N} {\frac{N}{2}\left( 1+ \frac{x}{\sqrt N} \right) } \oppropto_{N \to +\infty}  2^N \sqrt{ \frac{2}{\pi N}} e^{- x^2}
\label{binomialx}
\end{eqnarray}
one obtains
\begin{eqnarray}
 \sum_{k=1}^N k^{- \frac{1}{2}}   \binom {N} {k} \simeq   \frac{ 2^N }{\sqrt N } 
\label{sumbinom}
\end{eqnarray}
so that the criterion of Eq. \ref{yqmbltoy} becomes a bound for $\overline{  \vert J_{i_1,..,i_{\frac{N}{2}}} \vert }  $ in the thermodynamic limit
\begin{eqnarray}
 \overline{  \vert J_{i_1,..,i_{\frac{N}{2}}} \vert }  < \frac{\sqrt N } { 2^N }
\label{yqmbltoyleading}
\end{eqnarray}
Since the right-handside represents the scaling of the energy-level-spacing in the middle of the spectrum, the physical 
interpretation of this bound is thus very clear and natural :
 the MBL-phase is stable only if the coupling between two configurations having 
an extensive number $\frac{N}{2}$ of different spins is smaller than the level spacing $\Delta_N \propto \sqrt{N} 2^{-N}$.

\section{ Conclusion }

\label{conclusion}

In summary, the strong disorder perturbative expansion of the Local Integrals of Motion (Lioms) 
around the real-spin operators has been described to make the link with other properties of the Many-Body-Localized phase,
in particular the statistics of matrix elements of local operators in the eigenstate basis, and the memory in the dynamics of the local magnetizations. 
The lowest-order contribution to the magnetization memory was discussed for the random field XXZ chain, which is the most studied MBL model.
Finally, this approach has been used to analyze the MBL-transition
in the toy model considered previously in \cite{c_toy,c_pseudocriti} via other points of view.

As a final remark, it is interesting to mention the recent work on strong zero modes in the non-random XYZ chain  \cite{fendley}:
although it is not directly related to the problem of Many-Body-Localization in random models, 
the construction of an exact non-trivial operator that squares to the identity and whose commutator with the Hamiltonian is exponentially small
in the system size \cite{fendley} is nevertheless somewhat reminiscent of the notion of Lioms in random models.

\appendix

\section{ Non-perturbative Lioms for the simplest two-spin Hamiltonian  } 

\label{appendice}

Besides the perturbative framework described in the text for systems with a large number $N$ of spins,
 it is useful to see on the trivial case involving only $N=2$ spins
\begin{eqnarray}
H  = h_1 \sigma_1^z +h_2 \sigma_2^z+ \Delta \sigma_1^z \sigma_2^z + 2J (\sigma_1^+ \sigma_2^- +\sigma_1^- \sigma_2^+ )  
\label{htwos}
\end{eqnarray}
 how the finite unitary transformation $U=e^{\eta} $ based on the antihermitian generator $\eta$
\begin{eqnarray}
U=e^{\eta}= e^{\theta (\sigma_1^+ \sigma_2^- -\sigma_1^- \sigma_2^+ )  } 
= 1 + (\cos \theta -1) \frac{1-\sigma_1^z \sigma_2^z}{2} + \sin \theta (\sigma_1^+ \sigma_2^- -\sigma_1^- \sigma_2^+ )
\label{Utwos}
\end{eqnarray}
allows to obtain the pseudo-spins and the other observables described in the text.

\subsection { Lioms in terms of spins }

Eq \ref{Utwos} yields the pseudo-spins
\begin{eqnarray}
\tau_1^z && = U \sigma_1^z U^{\dagger} =\cos^2 \theta \sigma_1^z + \sin^2 \theta \sigma_2^z
 - 2 \cos \theta \sin \theta (\sigma_1^+ \sigma_2^- + \sigma_1^- \sigma_2^+ )
\nonumber \\
\tau_2^z && = U \sigma_2^z U^{\dagger} =\sin^2 \theta \sigma_1^z + \cos^2 \theta \sigma_2^z
 + 2 \cos \theta \sin \theta (\sigma_1^+ \sigma_2^- + \sigma_1^- \sigma_2^+ )
\label{tau2s}
\end{eqnarray}
and the ladder operators
\begin{eqnarray}
\tau_1^{\pm} && = U \sigma_1^{\pm} U^{\dagger} =\cos \theta \sigma_1^{\pm} + \sin \theta \sigma_1^z \sigma_2^{\pm} 
\nonumber \\
\tau_2^{\pm} && = U \sigma_2^{\pm} U^{\dagger} =- \sin \theta \sigma_1^{\pm} \sigma_2^z + \cos \theta \sigma_2^{\pm}
\label{ladder2s}
\end{eqnarray}
so that the hopping terms read for instance
\begin{eqnarray}
\tau_1^+ \tau_2^- && =\cos^2 \theta \sigma_1^+ \sigma_2^- -  \sin^2 \theta \sigma_1^- \sigma_2^+ + \cos \theta \sin \theta 
\frac{\sigma_1^z -\sigma_2^z}{2}  
\nonumber \\
\tau_1^- \tau_2^+ && =\cos^2 \theta \sigma_1^- \sigma_2^+ -  \sin^2 \theta \sigma_1^+ \sigma_2^- + \cos \theta \sin \theta 
\frac{\sigma_1^z -\sigma_2^z}{2}  
\label{hopping2s}
\end{eqnarray}

\subsection { Spins in terms of Lioms }

In terms of the pseudo-spins, the generator keeps the same form as a consequence of Eq. \ref{hopping2s}
\begin{eqnarray}
\eta = \theta (\sigma_1^+ \sigma_2^- -\sigma_1^- \sigma_2^+ )  = \theta (\tau_1^+ \tau_2^- -\tau_1^- \tau_2^+ )
\label{eta2s}
\end{eqnarray}
and leads to the expression of the real spins in terms of the pseudo-spins
\begin{eqnarray}
\sigma_1^z && = U^{\dagger}  \tau_1^z U = \cos^2 \theta \tau_1^z + \sin^2 \theta \tau_2^z
 + 2 \cos \theta \sin \theta (\tau_1^+ \tau_2^- + \tau_1^- \tau_2^+ )
\nonumber \\
\sigma_2^z && = U^{\dagger}  \tau_2^z U= \sin^2 \theta \tau_1^z + \cos^2 \theta \tau_2^z
 - 2 \cos \theta \sin \theta (\tau_1^+ \tau_2^- + \tau_1^- \tau_2^+ )
\end{eqnarray}

\subsection { Choice of the angle $\theta$ to diagonalize the Hamiltonian }

The Hamiltonian of Eq. \ref{htwos} reads in terms of the pseudo-spins
\begin{eqnarray}
H  && = h_1 \left( \cos^2 \theta \tau_1^z + \sin^2 \theta \tau_2^z
 +  \sin (2\theta) (\tau_1^+ \tau_2^- + \tau_1^- \tau_2^+ )\right) 
+h_2 \left( \sin^2 \theta \tau_1^z + \cos^2 \theta \tau_2^z
 -  \sin (2\theta )(\tau_1^+ \tau_2^- + \tau_1^- \tau_2^+ )\right) 
\nonumber \\ &&+ \Delta \tau_1^z \tau_2^z
 + 2J
 \left( \cos(2 \theta)  (\tau_1^+ \tau_2^- + \tau_1^- \tau_2^+)  - \cos \theta \sin \theta (\sigma_1^z -\sigma_2^z) 
  \right)  
\nonumber \\ &&
= \left( h_1 \cos^2 \theta+ h_2 \sin^2 \theta - J \sin(2 \theta) \right) \tau_1^z 
+  \left(  h_1 \sin^2 \theta+ h_2 \cos^2 \theta + J \sin(2 \theta) \right)  \tau_2^z+ \Delta \tau_1^z \tau_2^z
 \nonumber \\ &&+ ( (h_1-h_2) \sin (2\theta)+ 2 J  \cos(2 \theta)   ) (\tau_1^+ \tau_2^- + \tau_1^- \tau_2^+ )
\label{H2stau}
\end{eqnarray}
It is is thus diagonal in the $\tau^z$ basis when the hopping term vanishes, i.e. for the choice of the angle $\theta \in ] -\frac{\pi}{4},+\frac{\pi}{4} [$ satisfying
\begin{eqnarray}
\cos (2 \theta) && = \frac{1}{\sqrt{ 1+\left( \frac{ 2J}{-h_1+h_2} \right)^2 }}
\nonumber \\ 
\sin (2 \theta) && = \frac{ \frac{ 2J}{(-h_1+h_2)} }{\sqrt{ 1+\left( \frac{ 2J}{-h_1+h_2} \right)^2 }}
\label{sin2theta}
\end{eqnarray}
leading to
\begin{eqnarray}
H  && 
= \left( \frac{h_1+h_2}{2} +\frac{h_1-h_2}{2} \sqrt{ 1+\left( \frac{ 2J}{-h_1+h_2} \right)^2 } \right) \tau_1^z 
+\left( \frac{h_1+h_2}{2} - \frac{h_1-h_2}{2} \sqrt{ 1+\left( \frac{ 2J}{-h_1+h_2} \right)^2 } \right)     \tau_2^z+ \Delta \tau_1^z \tau_2^z
\nonumber 
\end{eqnarray}

\subsection{ Matrix elements of a single real spin operator in the eigenstate basis }

The matrix element of Eq \ref{mni}
\begin{eqnarray}
 < T_1 T_2 \vert \sigma_1^z \vert T_1' T_2'> =  (T_1\cos^2 \theta + T_2 \sin^2 \theta ) \delta_{T_1=T_1'} \delta_{T_2=T_2'} 
 +  \sin (2\theta) \delta_{T_1=-T_2} \delta_{T_1=-T_1'} \delta_{T_2=-T_2'} 
\label{mni2s}
\end{eqnarray}
yields the doubly stochastic matrix of Eq. \ref{qttprime}
\begin{eqnarray}
 Q_{T_1 T_2, T_1' T_2'} && \equiv   \vert  < T_1 T_2 \vert \sigma_1^z \vert T_1' T_2'>   \vert^2
\nonumber \\ && =
  (\cos^4 \theta + \sin^4 \theta + 2 T_1 T_2 \cos^2 \theta \sin^2 \theta ) \delta_{T_1=T_1'} \delta_{T_2=T_2'} 
 +  \sin^2 (2\theta) \delta_{T_1=-T_2} \delta_{T_1=-T_1'} \delta_{T_2=-T_2'} 
\label{qttprime2s}
\end{eqnarray}

\subsection{ Dynamics of the local magnetizations }

The magnetization of Eq. \ref{mpinfty} reads
\begin{eqnarray}
 m_{n=n_0=1}^{\infty} && =
 \frac{1}{4}  \sum_{T_1=\pm 1, T_2=\pm 1}  Q_{T_1 T_2, T_1 T_2} = \cos^4 \theta + \sin^4 \theta 
= ( \cos^2 \theta + \sin^2 \theta)^2 - 2 \cos^2 \theta \sin^2 \theta
\nonumber \\ && = 1- \frac{1}{2} \sin^2 (2 \theta )
\label{mpinfty2s}
\end{eqnarray}
or more explicitly using Eq \ref{sin2theta}
\begin{eqnarray}
 m_{n=n_0=1}^{\infty} && =
 1- 2 \frac{  J^2 }{(-h_1+h_2)^2+ 4 J^2 }
\label{mpinfty2sres}
\end{eqnarray}
When the two random fields $(h_1,h_2)$ are given, the correction is of second order in the coupling $J$
\begin{eqnarray}
 m_{n=n_0=1}^{\infty} && =
 1- 2 \frac{  J^2 }{(-h_1+h_2)^2 } +o(J^2)
\label{mpinfty2sresper}
\end{eqnarray}
but the averaged value over the two random fields gives a correction of order $\vert J \vert$ via the change of variable $h_2=h_1+ 2 \vert J \vert x$
\begin{eqnarray}
\overline{ m_{n=n_0=1}^{\infty} } && =
 1- \int dh_1 P(h_1) \int dh_2 P(h_2) 2 \frac{  J^2 }{(-h_1+h_2)^2+ 4 J^2 }
\nonumber \\
&& = 1-  \vert J \vert \int dh_1 P(h_1) \int dx P(h_1+ 2 \vert J \vert x)  \frac{  1 }{  (x^2+1) }
\nonumber \\
&& = 1-  \pi \vert J \vert \int dh_1 P^2 (h_1) + o( \vert J \vert)
\label{mpinfty2sresperav}
\end{eqnarray}

\section{ Magnetization memory in terms of the L\'evy sum of correlated variables }

\label{app_levy}

The Strong Disorder perturbative approach described in the text 
yields that the magnetization memory of Eq. \ref{mpinftynorm} reads
\begin{eqnarray}
 m_{n=n_0}^{\infty} = \frac{  1 } {  1+\Sigma   } 
\label{msigma}
\end{eqnarray}
in terms of the sum of $M$ positive terms 
\begin{eqnarray}
 \Sigma = \sum_{i=1}^M x_i
\label{sumlevy}
\end{eqnarray}
of the form
\begin{eqnarray}
 x_i =  \frac{J_i^2}{y_i^2}
\label{xiyi}
\end{eqnarray}
where $J_i$ are the fixed small perturbative off-diagonal couplings, while the variables $y_i$ depending on the random fields
are correlated random variables, such that the partial law $\rho_i(y_i)$ of $y_i$ has some finite weight $\rho_i(y_i=0)>0$ at the origin $y_i=0$.
As a consequence, the partial law $P_i(x_i)$ of $x_i$ of Eq. \ref{xiyi} displays the following L\'evy power-law tail for large $x_i$
\begin{eqnarray}
P_i(x_i) && = \int_{-\infty}^{+\infty} dy_i \rho_i(y_i) \delta\left( x_i -  \frac{J_i^2}{y_i^2} \right)
= \int_{-\infty}^{+\infty} dy_i \rho_i(y_i) \frac{ \delta\left( y_i -  \frac{J_i}{\sqrt{x_i} } \right) + \delta\left( y_i +  \frac{J_i}{\sqrt{x_i} } \right) }
{ 2 \frac{ x_i^{\frac{3}{2}} }{\vert J_i \vert} } 
\opsimeq_{x_i \to +\infty} \frac{\vert J_i \vert \rho_i(0)  }{x_i^{\frac{3}{2}}}
\label{Pxiyi}
\end{eqnarray}
and its Laplace transform displays the following singular behavior for small $p$
\begin{eqnarray}
\overline{  e^{-p x_i} } && = \int_0^{+\infty} dx_i P_i(x_i) e^{-p x_i} = 1-  \int_0^{+\infty} dx_i P_i(x_i) (1-e^{-p x_i})
=  1-  \int_0^{+\infty} \frac{du}{p}  P_i \left( \frac{u}{p} \right) (1-e^{-u} )
\nonumber \\
&& =  1- p^{\frac{1}{2}} \vert J_i \vert \rho_i(0) \int_0^{+\infty} du  \frac{  (1-e^{-u} )}{u^{\frac{3}{2}}} + o( p^{\frac{1}{2}} ) 
\nonumber \\
&& =  1- p^{\frac{1}{2}} \vert J_i \vert \rho_i(0) 2 \sqrt{\pi} + o( p^{\frac{1}{2}} ) 
\label{laplacex}
\end{eqnarray}

Since the average value of $x_i$ diverges
\begin{eqnarray}
\overline{  x_i } = \infty
\label{avdvx}
\end{eqnarray}
 the average value of the sum of Eq. \ref{sumlevy} also diverges
\begin{eqnarray}
\overline{  \Sigma } = \infty
\label{avdv}
\end{eqnarray}
To characterize the statistical properties of the sum $\Sigma$, one thus needs to evaluate the singular behavior 
of the Laplace transform of its probability distribution for small $p$ using Eq. \ref{laplacex}
\begin{eqnarray}
1- \overline{  e^{-p\Sigma} } && = 1- \overline{ \prod_{i=1}^M (1- (1-e^{-p x_i }) }
=  \sum_{i=1}^M \overline{   (1-e^{-p x_i }) } + o( p^{\frac{1}{2}} ) 
\nonumber \\
&&  = p^{\frac{1}{2}} \left(  \sum_{i=1}^M  \vert J_i \vert \rho_i(0) \right)  2 \sqrt{\pi} + o( p^{\frac{1}{2}} ) 
\label{laplace}
\end{eqnarray}

The disorder-averaged value of the magnetization memory of Eq. \ref{msigma} can be computed
from the Laplace transform of the probability distribution of $\Sigma$ via
\begin{eqnarray}
\overline{  m_{n=n_0}^{\infty} } = \overline{  \frac{  1 } {  1+\Sigma   } } = \int_0^{+\infty} dp e^{-p} \ \ \  \overline{ e^{-p \Sigma } }
\label{mpinftynormsl}
\end{eqnarray}
Eq \ref{laplace} then yields that the lowest order in the off-diagonal couplings $J_i$ reads
\begin{eqnarray}
\overline{ m_{n=n_0}^{\infty} } && =1 -   \int_0^{+\infty} dp e^{-p} (1- \overline{ e^{-p \Sigma } } )
\nonumber \\
&& = 1 -  \left(  \sum_{i=1}^M  \vert J_i \vert \rho_i(0) \right)  2 \sqrt{\pi} 
 \int_0^{+\infty} dp e^{-p} 
 p^{\frac{1}{2}} + o( \vert J_i \vert ) 
\nonumber \\
&& = 1 -  \left(  \sum_{i=1}^M  \vert J_i \vert \rho_i(0) \right)  \pi
+ o( \vert J_i \vert ) 
\label{mav}
\end{eqnarray}


\begin{thebibliography}{99}



\bibitem{revue_huse}
R. Nandkishore and D. A. Huse, Ann. Review of Cond. Mat. Phys. 6, 15 (2015).

\bibitem{revue_altman}
 E. Altman and R. Vosk, Ann. Review of Cond. Mat. Phys. 6, 383 (2015).

\bibitem{revue_vasseur}
S. A. Parameswaran, A. C. Potter and R. Vasseur, Annalen der Physik , 1600302 (2017).

\bibitem{revue_imbrie}
J. Z. Imbrie, V. Ros and A. Scardicchio, Annalen der Physik, 1600278 (2017)

\bibitem{revue_rademaker}
L. Rademaker, M. Ortuno and A.M. Somoza,  Annalen der Physik 1600322 (2017)

\bibitem{review_mblergo}
D. J. Luitz, Y. Bar Lev, Annalen der Physik  1600350 (2017)

\bibitem{review_prelovsek}
P. Prelovsek, M. Mierzejewski, O. Barisic, J. Herbrych, Annalen der Physik  1600362 (2017)

\bibitem{review_rare}
K. Agarwal {\it et al}, Annalen der Physik  1600326 (2017)


\bibitem{emergent_swingle}
B. Swingle, arxiv:1307.0507.

\bibitem{emergent_serbyn}
M. Serbyn, Z. Papic and D.A. Abanin, Phys. Rev. Lett. 111, 127201 (2013).

\bibitem{emergent_huse}
D.A. Huse, R. Nandkishore and V. Oganesyan, Phys. Rev. B 90, 174202 (2014).

\bibitem{emergent_ent}
A. Nanduri, H. Kim and D.A. Huse, Phys. Rev. B 90, 064201 (2014).

\bibitem{imbrie}
J. Z. Imbrie, J. Stat. Phys. 163, 998 (2016).

\bibitem{serbyn_quench}
M. Serbyn, Z. Papic and D.A. Abanin, Phys. Rev. B 90, 174302 (2014).

\bibitem{emergent_vidal}
A. Chandran, I.H. Kim, G. Vidal and  D.A. Abanin, Phys. Rev. B 91, 085425 (2015).

\bibitem{emergent_ros}
V. Ros, M. M\"uller and A. Scardicchio, Nucl. Phys. B 891, 420 (2015).

\bibitem{emergent_rademaker}
L. Rademaker and M. Ortuno, Phys. Rev. Lett. 116, 010404 (2016).

\bibitem{serbyn_powerlawent}
M. Serbyn, A. A. Michailidis, D. A. Abanin, Z. Papic, Phys. Rev. Lett. 117, 160601 (2016).

\bibitem{c_emergent}
C. Monthus, J. Stat. Mech. (2016) 033101.

\bibitem{ros_remanent}
V. Ros and M. Mueller,  	Phys. Rev. Lett. 118, 237202 (2017)

\bibitem{wortis}
R Wortis and Malcolm P Kennett, J. Phys.: Condens. Matter 29, 405602 (2017)

\bibitem{c_toy}
C. Monthus, Entropy 18, 122 (2016).

\bibitem{c_pseudocriti}
C. Monthus, J. Stat. Mech. (2016) 123303

\bibitem{serbyn_criterion}
M. Serbyn, Z. Papic and D.A. Abanin, Phys. Rev. X 5, 041047 (2015).




\bibitem{keating}
J. P. Keating, N. Linden, H. J. Wells, Comm. Math. Phys. 338, 81 (2015).

\bibitem{c_toda}
C. Monthus, J. Phys. A: Math. Theor. 49 305002 (2016).

\bibitem{serbyn_dyson}
M. Serbyn and J.E. Moore, Phys. Rev. B 93, 041424(R) (2016).


\bibitem{c_dysonBM}
C. Monthus,  J. Stat. Mech. (2016) 033113.

\bibitem{c_mblstrongmultif}
C. Monthus, J. Stat. Mech. (2016) 073301.



\bibitem{deutsch}
J.M. Deutsch, Phys. Rev. A 43, 2046 (1991).

\bibitem{srednicki}
M. Srednicki, Phys. Rev. E 50, 888 (1994).

\bibitem{nature}
M. Rigol, V. Dunjko and M. Olshanii, Nature 452, 854 (2008)

\bibitem{mite}
S. Goldstein, D.A. Huse, J.L. Lebowitz and R. Tumulka, Phys. Rev. Lett. 115,
100402 (2015).

\bibitem{rigol}
L. D'Alessio, Y. Kafri, A. Polkovnikov and M. Rigol, Adv. Phys. 65, 239 (2016).


\bibitem{us_stronglevy}
C, Monthus and T. Garel, J. Stat. Mech. (2010) P09015.


\bibitem{kjall}
J. A. Kj\"all, J. H. Bardarson and F. Pollmann, Phys. Rev. Lett. 113, 107204 (2014).

\bibitem{alet}
D. J. Luitz, N. Laflorencie and F. Alet, Phys. Rev. B 91, 081103 (2015).

\bibitem{alet_dyn}
D. J. Luitz, N. Laflorencie and F. Alet, Phys. Rev. B 93, 060201 (2016).


\bibitem{luitz_tail}
D. Luitz, Phys. Rev. B 93, 134201 (2016).

\bibitem{badarson_signa}
R. Singh, J. H. Bardarson and F. Pollmann, New J. Phys. 18, 023046 (2016).

\bibitem{auerbach}
I. Khait, S. Gazit, N. Y. Yao, A. Auerbach, Phys. Rev. B 93, 224205 (2016).

\bibitem{znidaric_dephasing}
M. V. Medvedyeva, T. Prosen, M. Znidaric, Phys. Rev. B 93, 094205 (2016).

\bibitem{prelo_dyn}
O. S. Barisic, J. Kokalj, I. Balog and P. Prelovsek, Phys. Rev. B 94, 045126 (2016).

\bibitem{znidaric_lindblad}
M. Znidaric, A. Scardicchio, V. K. Varma, Phys. Rev. Lett. 117, 040601 (2016)

\bibitem{luitz_bimodal}
X. Yu, D. J. Luitz, B. K. Clark, Phys. Rev. B 94, 184202 (2016)

\bibitem{garcia}
C.L. Bertrand and A.M. Garcia-Garcia, Phys. Rev. B 94, 144201 (2016).

\bibitem{luitz_anomalous}
D. J. Luitz and  Y. Bar Lev, Phys. Rev. Lett. 117, 170404 (2016).

\bibitem{luitz_operator}
T. Zhou and D. J. Luitz, Phys. Rev. B 95, 094206 (2017)

\bibitem{luitz_information}
D. J. Luitz and Y. Bar Lev, Phys. Rev. B 96, 020406 (2017)



\bibitem{fendley}
P. Fendley, J. Phys. A Math. Theor. 49, 30LT01 (2016).

 \end{thebibliography}
\end{document}